\documentclass[preprint,showpacs,nofootinbib,prd,aps]{revtex4-1}
\usepackage[utf8]{inputenc}
\usepackage{graphicx,amstext,amssymb}

\begin{document}
\title{Description of bulk observables in Au+Au collisions at top RHIC energy in the integrated HydroKinetic Model}

\author{M.~D.~Adzhymambetov$^{1}$}
\author{V.~M.~Shapoval$^{1}$}
\author{Yu.~M.~Sinyukov$^{1}$}

\affiliation{$^1$Bogolyubov Institute for Theoretical Physics, 03143 Kiev, Ukraine}

\begin{abstract}
The results on the main bulk observables obtained in the simulations within the integrated hydrokinetic model (iHKM) of  Au+Au collisions at the RHIC energy 
$\sqrt{s_{NN}}=200$~GeV  are presented along with the corresponding experimental data
from the STAR and the PHENIX collaborations. The simulations include all the stages of the collision process: formation of the initial state, its gradual thermalization and hydrodynamization, viscous relativistic hydro-evolution, system's hadronization and particlization, and, finally, an expansion of the interacting hadron-resonance gas. The model gives a satisfactory description of charged-particle multiplicities,
particle number ratios, transverse momentum spectra for pions, kaons, protons and antiprotons,  charged-particle $v_2$~coefficients, 
and femtoscopy radii at all collision centralities. It is demonstrated how one can estimate the times of the pion and kaon maximal emission from the femto-scales.   

\end{abstract}

\maketitle


\section{Introduction}
The comprehensive study of ultrarelativistic heavy ion collisions allow researchers gradually, step by step reveal 
new properties of rather interesting and unusual form of matter, created in these processes, and construct
more and more clear and full picture of evolution of such super dense and super hot systems.
As it became clear after the thorough analysis of bulk observables at RHIC and LHC, such as particle multiplicities,
transverse momentum spectra, and femtoscopy scales, the strongly interacting quark-gluon matter, formed in a collision
at high energy, at some stage of its evolution undergoes collective expansion and behaves like a nearly thermalized, (quasi)macroscopic system.
This fact justified the application of hydrodynamical and statistical mechanics approximations for the theoretical description of this stage.
However, the pre-thermal dynamics, leading to the system's thermalization, as well as the ``afterburner'' stage of its evolution,
also play an important role in the formation of final observables. That is why a realistic model, allowing to successfully describe
and predict various experimental data and helping to understand the reasons and mechanisms for the specific experiment results,
should be complex and include an adequate simulation of all the stages of the collision process.

In this work, we present the results of our study, devoted to the description of different bulk observables in Au+Au collisions at the 
RHIC energy $\sqrt{s_{NN}}=200$~GeV within such a decent model --- the integrated hydrokinetic model~\cite{ihkm} (iHKM).

Despite the experiments at the top RHIC energy were performed quite a long time ago, and the most recent results concern 
heavy ion collisions at the LHC, the datasets, collected at RHIC, are still in use and still are of interest for the analysis,
in particular, for the studies dealing with kaon femtoscopy~\cite{Grigory}. Additionally, although the time has passed since the 
first papers, presenting the results of certain measurements at RHIC (e.g., two-pion femtoscopy), were published, 
the STAR and PHENIX collaborations continue to issue new articles, containing results on the same topic, but with increased 
accuracy, in a wider region, with new cuts applied, etc. This fact also motivates one not to forget about the RHIC data.
  
Previously, the Au+Au collisions at the top RHIC energy were successfully simulated in the hydrokinetic model~\cite{hkm,uniform,phenixfemto}, 
the model-predecessor of the modern, more developed iHKM, which proved to be good in describing observables at the LHC energies~\cite{ihkm,ratios,lhc502}.
Here we aim to adjust the iHKM to the description of yields, $p_T$ spectra, interferometry radii, etc. at RHIC and see what differences in the model 
parameters and tuning will it require as compared to the LHC case.

\section{Model description}
In iHKM the process of the evolution of the system, formed in the relativistic nuclear collision starts with the pre-thermal stage,
which simulates the process of gradual transformation of the initially not thermalized system to a nearly thermal one, close to
local thermal and chemical equilibrium, that can be further described using viscous hydrodynamics approximation.
At this stage an energy-momentum transport approach in the relaxation time approximation is utilized (see \cite{ihkm, ihkm2} for details).

The initial distribution of energy density in the transverse plane for the pre-thermal stage is chosen to be a linear combination of wounded nucleons and 
binary collision contributions in GLISSANDO~\cite{gliss} Glauber Monte Carlo model:
\begin{equation}\label{eq1}
\epsilon(b,\textbf{r}_{T})=\epsilon_{0}(\tau_0)
\frac{(1-\alpha)N_w(b,\textbf{r}_{T})/2+\alpha N_{bin}(b,\textbf{r}_{T})}{(1-\alpha)N_w(b=0,\textbf{r}_{T}=0)/2+\alpha N_{bin}(b=0,\textbf{r}_{T}=0)},
\end{equation}
where the parameters $\alpha$ (defining the proportion between the two contributions to $\epsilon(b,\textbf{r}_{T})$) and 
$\epsilon_{0}(\tau_0)$ (defining the maximal initial energy density in the center of the system for the most central collisions) 
are adjusted to provide the best fit to experimental dependence of mean charged particle multiplicity in the pseudorapidity region $|\eta|<0.5$ on 
centrality, and the value of the starting time $\tau_0$ ensures the best description of pion $p_T$ spectrum slope in the most central events.
For the current study we obtained $\alpha=0.18$ and $\epsilon_{0}=235$~GeV/fm$^3$ at $\tau_0=0.1$~fm/$c$. At the LHC energies, the coefficient 
$\alpha=0.24$, and at the same initial time $\tau_0=0.1$~fm/$c$ one has $\epsilon_{0}=679$~GeV/fm$^3$ for $\sqrt{s_{NN}}=2.76$ TeV~\cite{ihkm}, 
and $\epsilon_{0}=1067$~GeV/fm$^3$ for $\sqrt{s_{NN}}=5.02$~TeV~\cite{lhc502}. The other parameters, such as the initial momentum anisotropy, 
viscosity-to-entropy ratios, relaxation and thermalization times, are the same at all the mentioned collision energies at RHIC and LHC and 
at the same Laine-Schroeder equation of state~\cite{Laine} (the latter is only corrected to take into account a small chemical potential at 
the top RHIC energy as discussed below).   

As for the initial momentum distribution for the pre-equilibrium dynamics, it is taken in the following ``Color-Glass-Condensate-like'' form:
\begin{eqnarray}
f_0(p)=g \exp\left(-\sqrt{\frac{(p\cdot U)^2-(p\cdot
V)^2}{\lambda_{\perp}^2}+\frac{(p\cdot
V)^2}{\lambda_{\parallel}^2}}\right),
\label{anis1}
\end{eqnarray}
where $U^{\mu}=( \cosh\eta, 0, 0, \sinh\eta)$, $V^{\mu}=(\sinh\eta, 0, 0,\cosh\eta)$, $\eta$ is space-time rapidity, 
and initial momentum anisotropy $\Lambda=\frac{\lambda_{\perp}}{\lambda_{\parallel}}=100$~\cite{ihkm}.

After the pre-thermal evolution in iHKM follows the viscous hydrodynamics stage, realized within the Israel-Stewart formalism.
Here we describe the collisions at RHIC, so, in contrast to the LHC case, we need to account for a small, but still non-zero chemical potential 
(baryon and strange) in the equation of state (EoS) for the quark-gluon phase. In order to do this, we modify the EoS at zero chemical potential
according to~\cite{eos}:
\begin{equation}\label{eos}
\frac{p(T,\mu_B,\mu_S)}{T^4}=\frac{p(T,0,0)}{T^4}+\frac{1}{2}\frac{\chi_B}{T^2	} {\left( \frac{\mu_B}{T}\right) }^2 +\frac{1}{2}\frac{\chi_S}{T^2	} {\left( \frac{\mu_S}{T}\right) }^2,
\end{equation}
where $p(T,0,0)$ is the Laine-Schroeder~\cite{Laine} equation of state at zero chemical potential and
\begin{equation}
\frac{\chi_i}{T^2}=\frac{1}{VT^3}\frac{\partial^2lnZ}{\partial(\mu_i/T)^2}, \ \ \  i=B,S.
\end{equation}
Here the ``mixed'' terms with $\chi_{BS}$, as well as the terms proportional to electric chemical potential $\mu_E$ are neglected due to their smallness. 
In this paper we take 
$\mu_B=21$~MeV at the hadronization/particlization hypersurface $T=165$~MeV for the best description of the ratio of proton yield to that of antiproton 
for the centrality $c=0-5\%$. And the value of $\mu_S=5$~MeV is obtained using the variation method as satisfying the condition of zero strangeness 
on the hadronization (particlization) hypersurface:
\begin{equation}
\left. S\right|_{\sigma_{p}}=0, \ \ \ S=\sum_{i}(N(i)-\overline{N}(i))\mu_{S,i}, \ \ 
\end{equation}
where the sum is taken over all particle species, $N(i)$ is the number of particles of the $i$th sort, $\overline{N}(i)$ is the number of 
corresponding antiparticles on the hadronization hypersurface, and $\mu_{S,i}$ is the strange chemical potential of
the particle species $i$. The evolution equations with chemical potential depending on $T$ are considered in the same way as in Refs.~\cite{hkm,uniform}. 

We assume that at the top RHIC energy the strange quarks do not have enough time to reach the chemical equilibrium
in {\it non-central collisions}, so that the kaon spectra get down. The same concerns 
proton spectra, since about a half of produced protons come from the decays of strange resonances (such as $\Lambda$, $\Sigma$, $\Xi$).
To take this into account we introduce an effective downscaling factor $\gamma_S(\tau_p)$, 
depending on the characteristic particlization time $\tau_p$ for each given centrality. This time is calculated in iHKM.
Each hadron yield is multiplied by $\gamma_S^{S_i}$, where $S_i$ is the strangeness of the hadron species $i$.
We assume the dependence
\begin{equation} 
\gamma_S(\tau_p) = A \exp(-b/\tau_p),
\label{gammas} 
\end{equation}
with $A=1.1$ and $b=0.8$~fm/$c$.
This choice guarantees $\gamma_S=1$ for the most central events and a good description of kaon and proton spectra,
together with $K/\pi$ ratio.
Table~\ref{gamma} shows the gamma factors, obtained for different centralities based on the corresponding particlization times.

\begin{table}
	
	\centering
	\begin{tabular}{|c|c|c|}
		\hline
		\ centrality [\%] \ & \ \ $\tau_{p}$ [fm/$c$] \ \ & \ \ \ \ \ $\gamma_{S}$ \ \ \ \ \ \\
		\hline
		0-10  & 7.55 & 0.989 \\
		\hline
		10-20 & 6.50 & 0.973 \\
		\hline
		20-30 & 5.60 & 0.953 \\
		\hline
		30-40 & 4.85 & 0.933 \\
		\hline
		40-50 & 4.20 & 0.909 \\
		\hline
		50-60 & 3.60 & 0.881 \\
		\hline
		60-70 & 3.00 & 0.843 \\
		\hline
		70-80 & 2.30 & 0.777 \\
		\hline
	\end{tabular}

	\caption {The values of $\gamma_{S}$ for different centrality classes.}
	\label{gamma}
\end{table}

After the hydrodynamic stage in iHKM we have the particlization stage, when we switch from the description of the 
system's evolution in terms of continuous medium to the description in terms of particles. In the current analysis this switching 
is performed at the hadronization hypersurface $T=165$~MeV. After that particles pass the stage of the hadronic cascade, realized 
in iHKM within the UrQMD model~\cite{urqmd}, which simulates resonance decays, as well as numerous elastic and inelastic hadron scatterings, 
taking place in expanding and initially dense hadronic gas. 


\section{Results and discussion}
Having adjusted the model parameters and run enough simulations to collect good statistics, 
we can immediately calculate a great variety of observables, basing on the model output.

\begin{figure}
	\centering
	\includegraphics[width=0.75\textwidth]{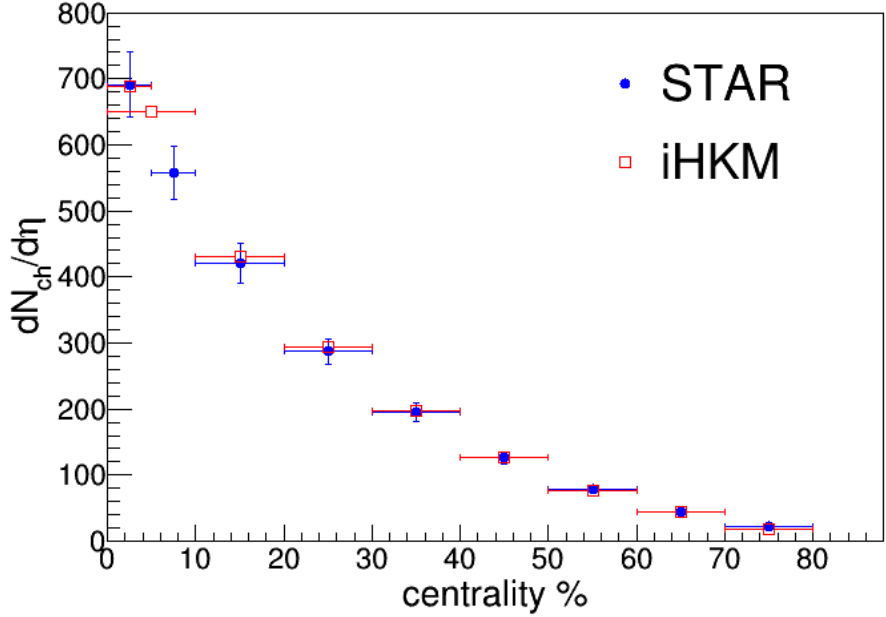}	
	\caption{The iHKM results on the mean charged particle density $\langle dN_\mathrm{ch}/d\eta\rangle$ at $|\eta|<0.5$ for different centrality classes,
	compared with the STAR data~\cite{multiplicity}.}
\label{mult1}
\end{figure}

In Fig.~\ref{mult1} and Table~\ref{mult} one can see the iHKM results on the mean multiplicity of all charged particles 
$\langle dN_\mathrm{ch}/d\eta\rangle$
in pseudorapidity range $|\eta|<0.5$ for eight centrality classes --- 0-5\%, 10-20\%, 20-30\%, 30-40\%, 40-50\%, 50-60\%, 60-70\%, 70-80\% --- 
in comparison with the STAR experimental data~\cite{multiplicity}. The model results are in good agreement with the experiment.

\begin{table}
	\begin{tabular}{|c|c|c|}
		\hline
		\ \ \ 	centrality [\%]\ \ \ \ &\ \ \  $\langle dN_\mathrm{ch}/d\eta\rangle$ in iHKM \ \ \ & \ \ \ \  $\langle dN_\mathrm{ch}/d\eta\rangle$ from STAR \ \ \ \   \\
		\hline
		0-5	  & 688 & $691 \pm 49$ \\
		\hline
		10-20 & 431 & $421 \pm 30$\\
		\hline
		20-30 & 294 & $287 \pm 20$\\
		\hline
		30-40 & 197 & $195 \pm 14$\\
		\hline
		40-50 & 127 & $126 \pm 9$\\
		\hline
		50-60 & 77  & $78 \pm 6$\\
		\hline
		60-70 & 45  & $45 \pm 3$\\
		\hline
		70-80 & 19  & $22 \pm 2$\\
		\hline
	\end{tabular}
	\caption {Mean charged particle density $\langle dN_\mathrm{ch}/d\eta\rangle$ at $|\eta|<0.5$ for different centrality classes. }
\label{mult}	
\end{table}

\begin{figure}
	\centering
	\includegraphics[width=0.75\textwidth]{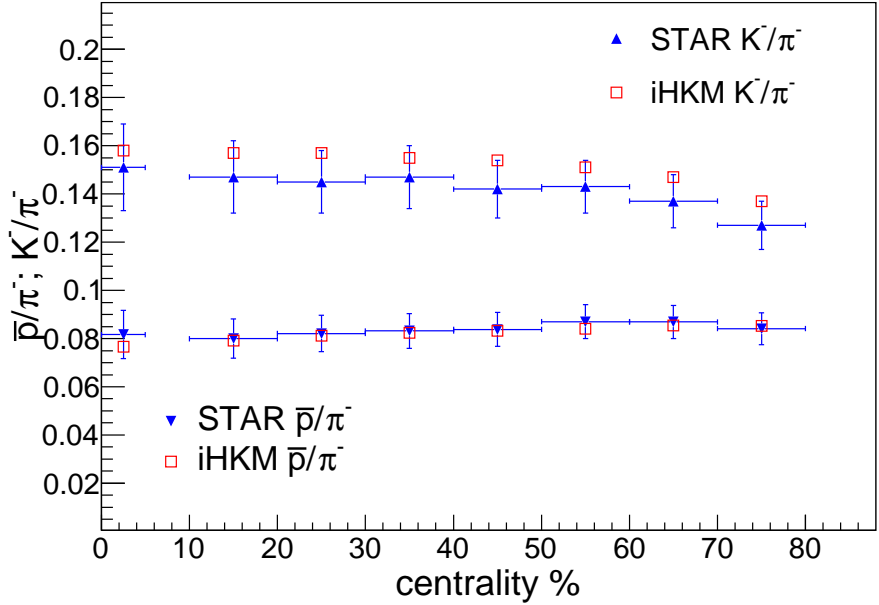}	
	\caption{The ratios of $K^{-}$ (upper points) and $\bar{p}$ (lower points) yields to that of $\pi^{-}$, obtained in iHKM in comparison with
	the STAR data~\cite{star}. Particles with $|y|<0.1$ were selected for the analysis.}
\label{pnr}
\end{figure}

In Fig.~\ref{pnr} we present our results on $K^{-}/\pi^{-}$ and $\bar{p}/\pi^{-}$ particle number ratios for different 
centrality classes in comparison with the experiment~\cite{star}.
The hadrons with rapidity from the range $|y|<0.1$ were selected to build all the ratios. As one can see from the figure, the iHKM describes
well both ratios at all centralities within the errors, however $K/\pi$ model points go through the upper edges of the experimental error bars. 

\begin{figure}
\includegraphics[width=0.75\textwidth]{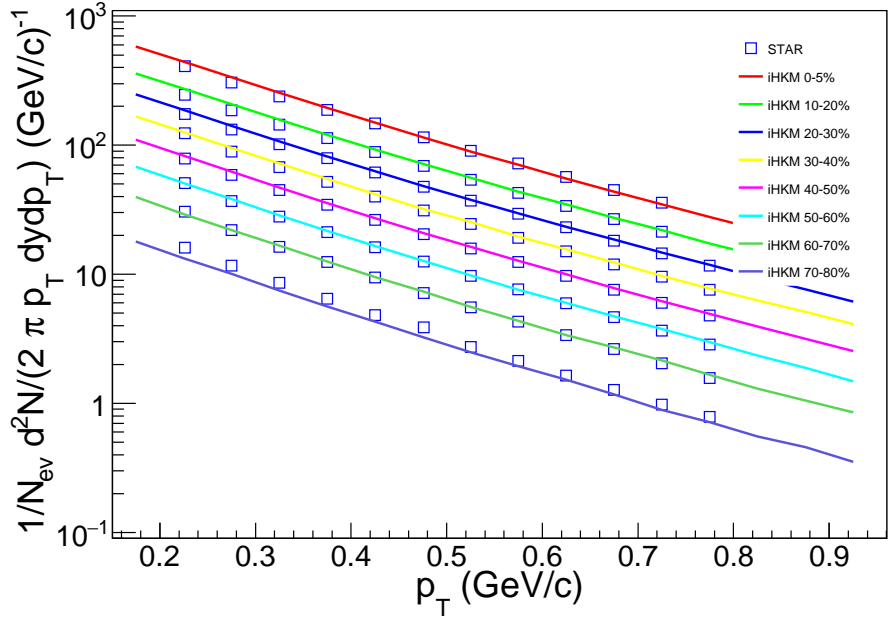}
\caption{The $p_T$ spectra of $\pi^{-}$ in comparison with the STAR data~\cite{star} at midrapidity, $|y|<0.1$.}
\label{pionspctr}
\end{figure}

\begin{figure}
\begin{center}
\includegraphics[width=0.75\textwidth]{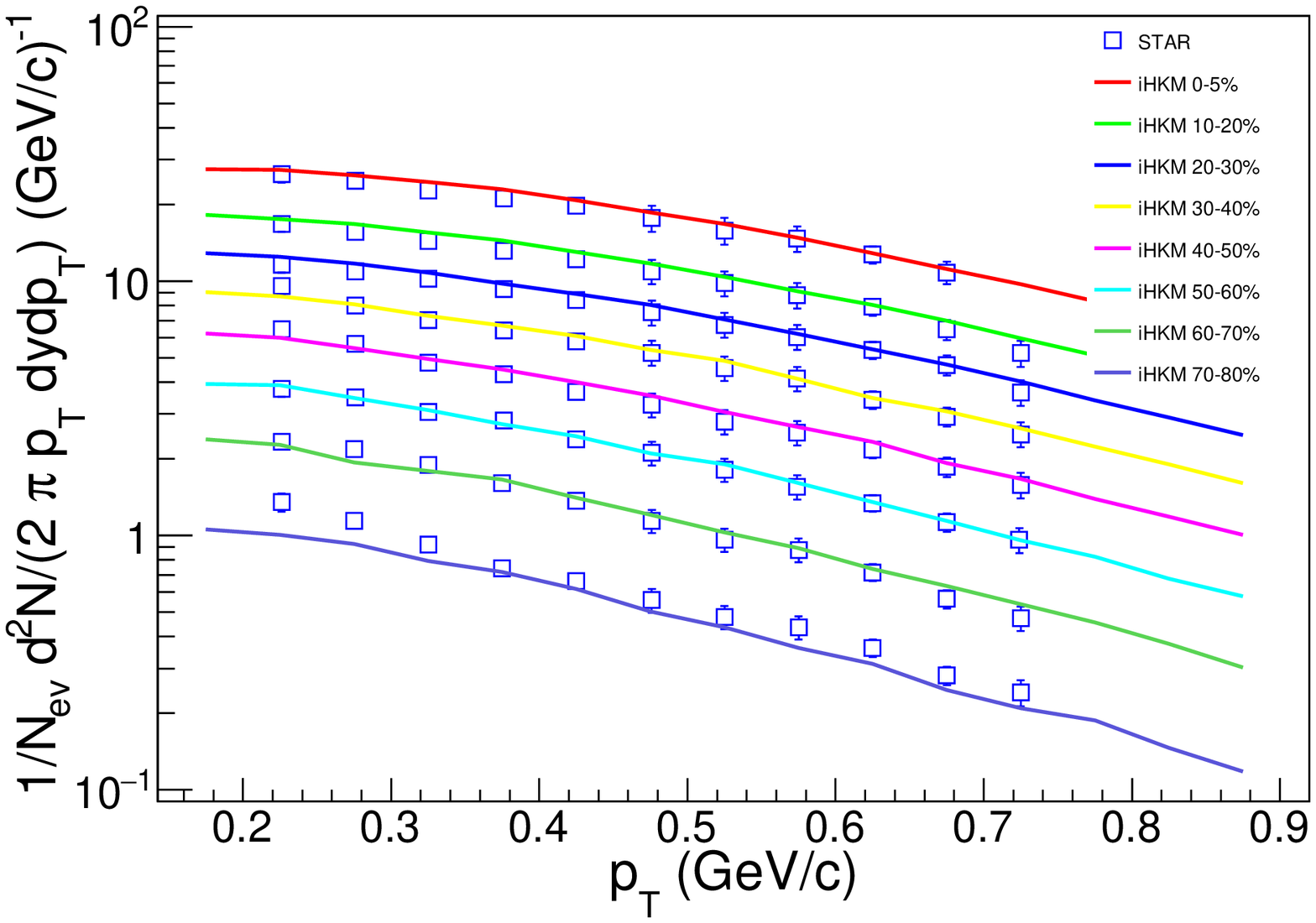}	
\caption{The $p_T$ spectra of $K^{-}$ in comparison with the STAR data~\cite{star} at midrapidity, $|y|<0.1$.}
\label{kaonspctr}
\end{center}
\end{figure}

\begin{figure}
\begin{center}
\includegraphics[width=0.75\textwidth]{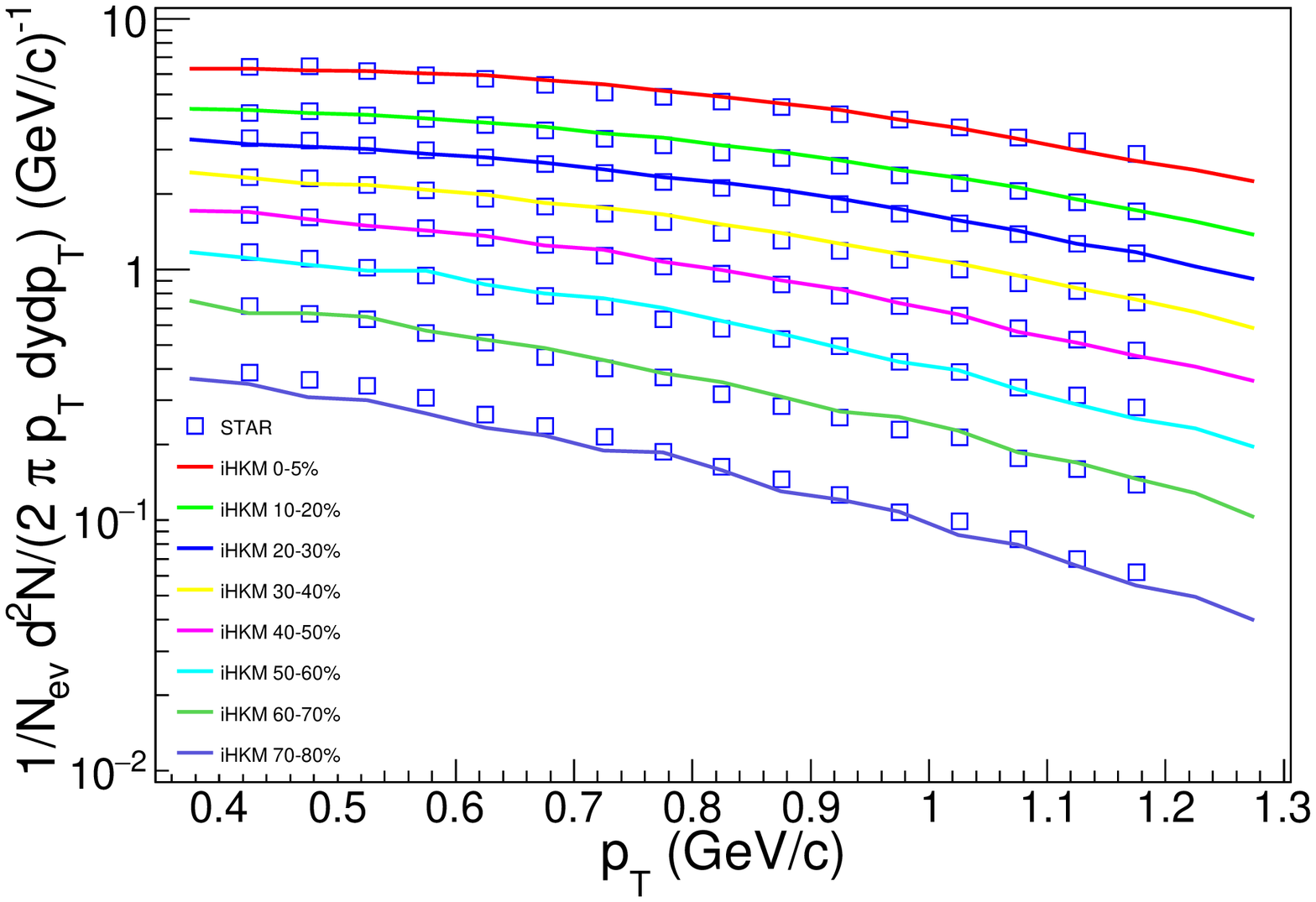}
\caption{The proton $p_T$ spectra in comparison with the STAR data~\cite{star} at midrapidity, $|y|<0.1$.}
\label{protonspctr}
\end{center}
\end{figure}

\begin{figure}
\begin{center}
\includegraphics[width=0.7\textwidth]{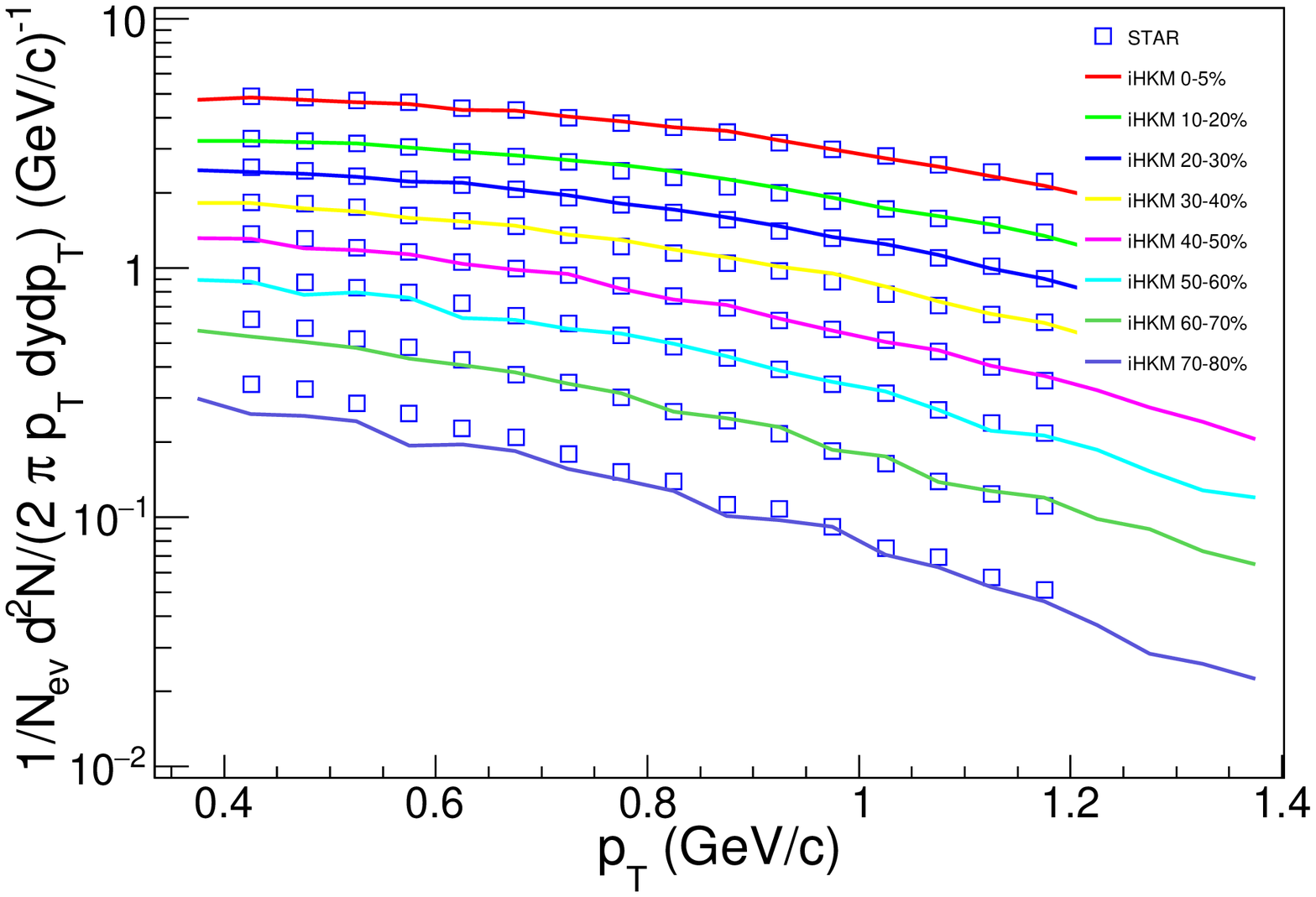}
\caption{The antiproton $p_T$ spectra in comparison with the STAR data~\cite{star} at midrapidity, $|y|<0.1$.}
\label{aprotonspctr}
\end{center}
\end{figure}

Figures~\ref{pionspctr}--\ref{aprotonspctr} demonstrate the iHKM description of transverse momentum spectra for negatively charged pions and kaons,
as well as for protons and antiprotons at different centralities in the rapidity range $|y|<0.1$. The simulation results are compared with
the STAR experimental data~\cite{star}. One can say that the model reproduces the measured spectra quite well for all the mentioned particle species 
and all the centrality classes. A slight deviations of the iHKM lines from the experimental points can be noticed in $K$ and $\bar{p}$ spectra 
at low $p_T$ for the very peripheral collisions with $c=70-80\%$ only. Note however, that such a good agreement with data is achieved here using a
``strangeness suppression'' $\gamma_S$ factor (\ref{gammas}), which helps to lower the ``raw'' model kaon and (anti)proton spectra for non-central events.

In the next Fig.~\ref{graf:v2} for three centrality classes, $c=10-20\%$, $c=20-30\%$, and $c=30-40\%$, we show
the iHKM results on $p_T$-dependence of the elliptic flow, or $v_2$ coefficients, characterizing the anisotropy of the 
all-charged-particles transverse momentum spectra. The model lines go in agreement with data for not very high $p_T$.

\begin{figure}
\begin{center}
\includegraphics[bb=0 0 567 409,width=0.7\textwidth]{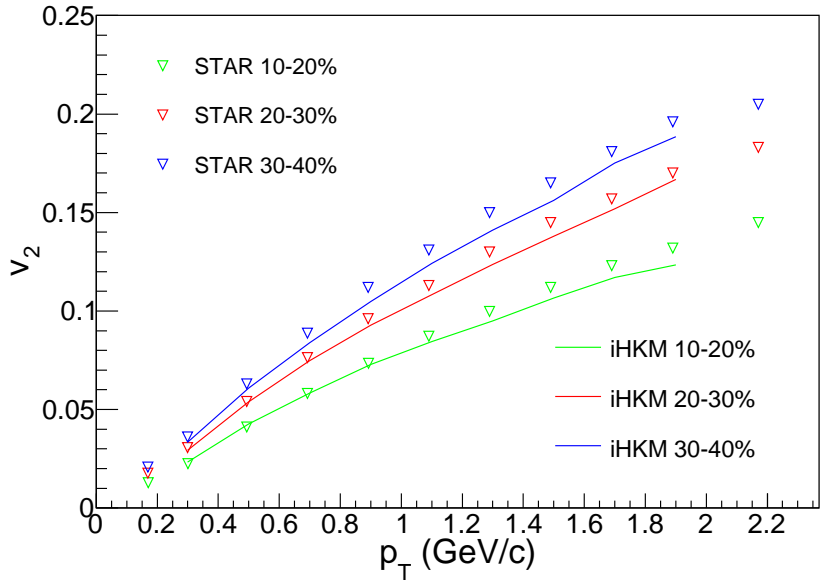}
\caption{The $v_2$ coefficients vs. $p_T$ for all charged particles, calculated in iHKM (lines), together with the STAR data~\cite{v2star} (triangles).
The results for the three centralities are shown: $c=10-20\%$, $c=20-30\%$, and $c=30-40\%$.}
\label{graf:v2}
\end{center}
\end{figure}

\begin{figure}
\begin{center}
\includegraphics[width=0.85\textwidth]{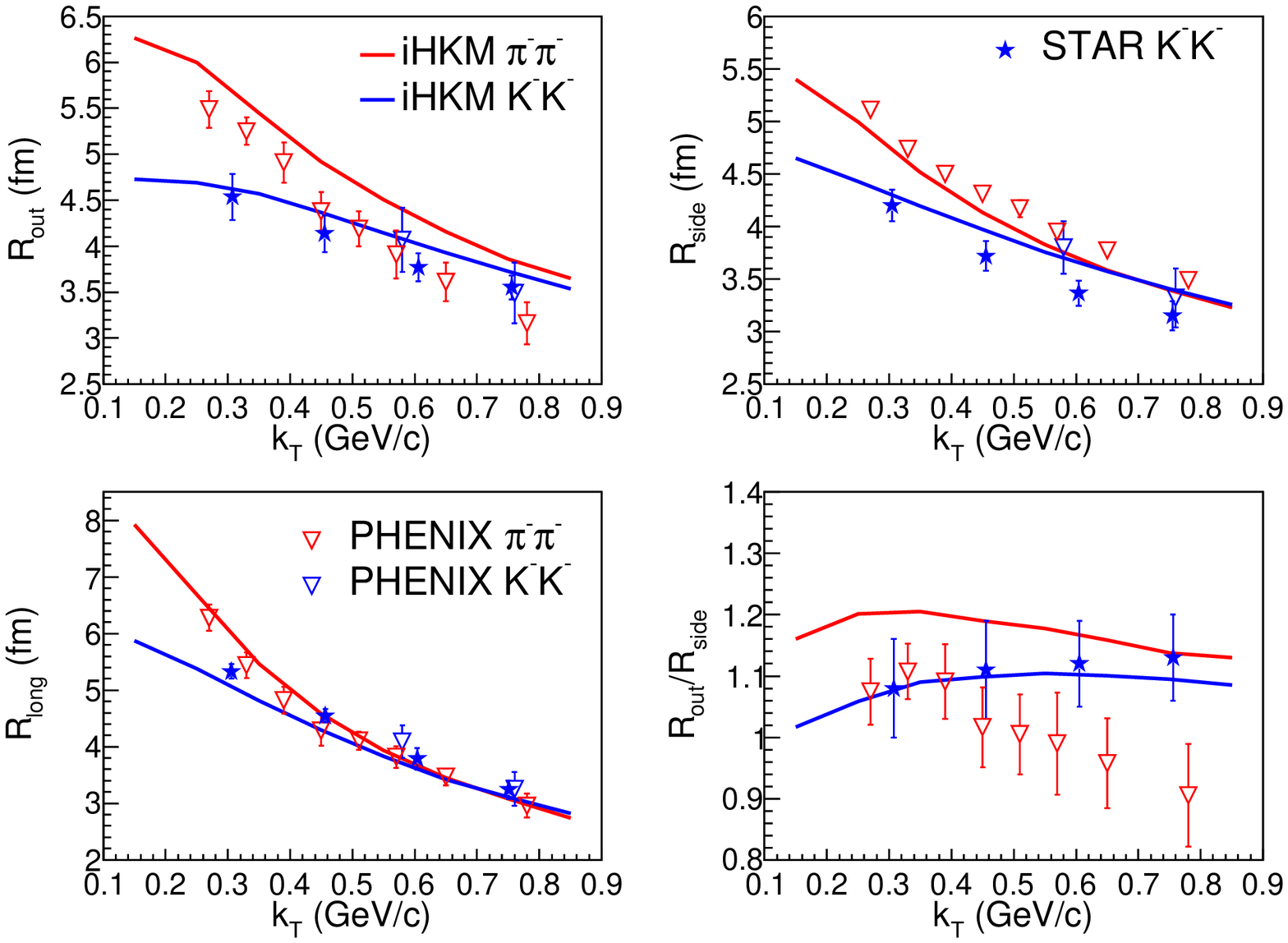}
\caption{The iHKM results on $\pi^{-}\pi^{-}$ and $K^{-}K^{-}$ femtoscopy scales vs. pair $k_T$ (lines) in comparison with the experimental data 
from PHENIX~\cite{phenixfemto} (triangles) and STAR~\cite{pionfemto,Grigory} (stars) for the events from the centrality class $c=0-10\%$. 
Red color is related to pions, blue color is related to kaons.}
\label{femto_0-10}
\end{center}
\end{figure}

\begin{figure}
\begin{center}
\includegraphics[width=0.85\textwidth]{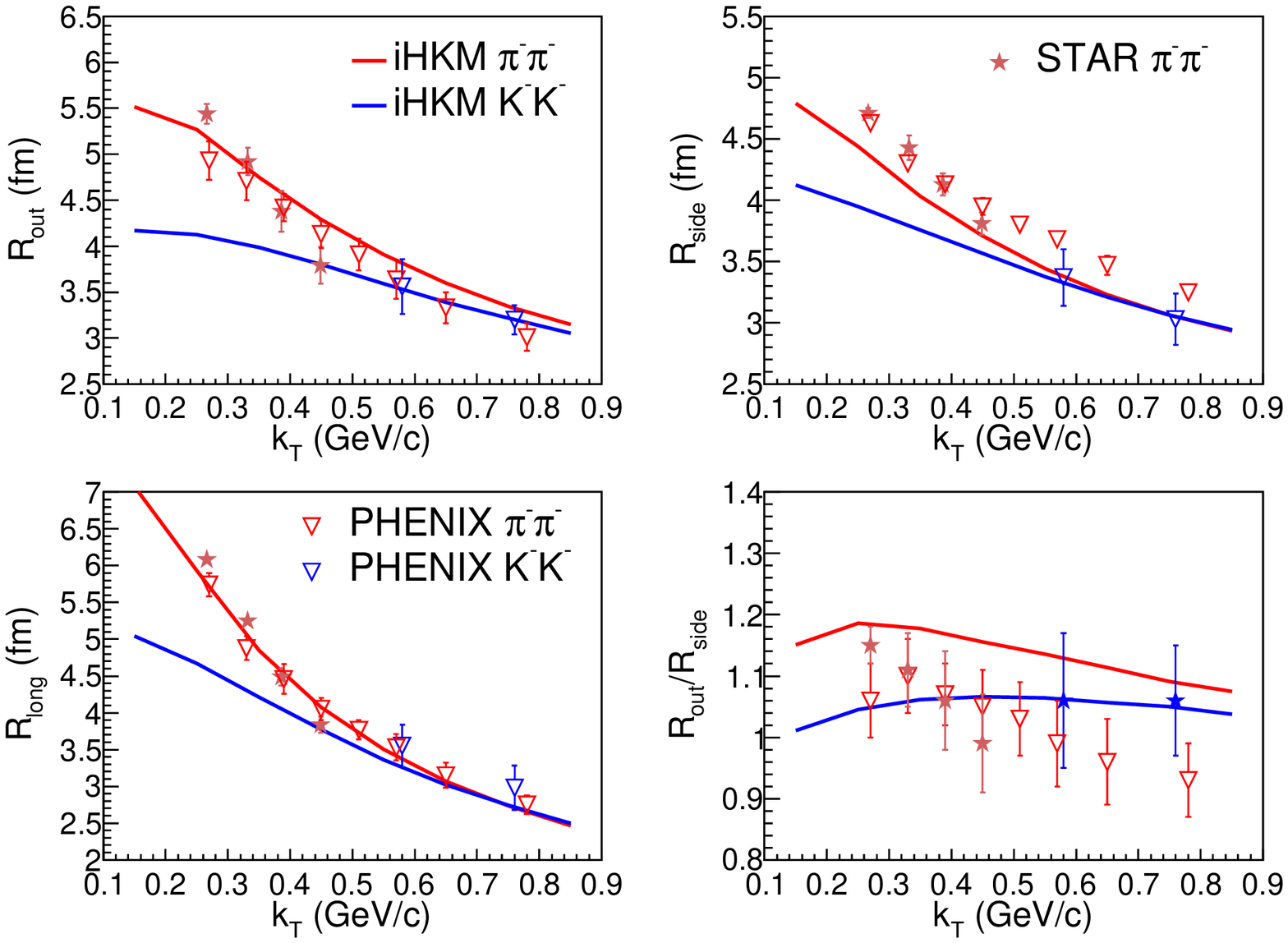}
\caption{The same as in Fig.~\ref{femto_0-10} for the centrality class $c=10-20\%$.}
\label{femto_10-20}
\end{center}
\end{figure}

\begin{figure}
\begin{center}
\includegraphics[width=0.85\textwidth]{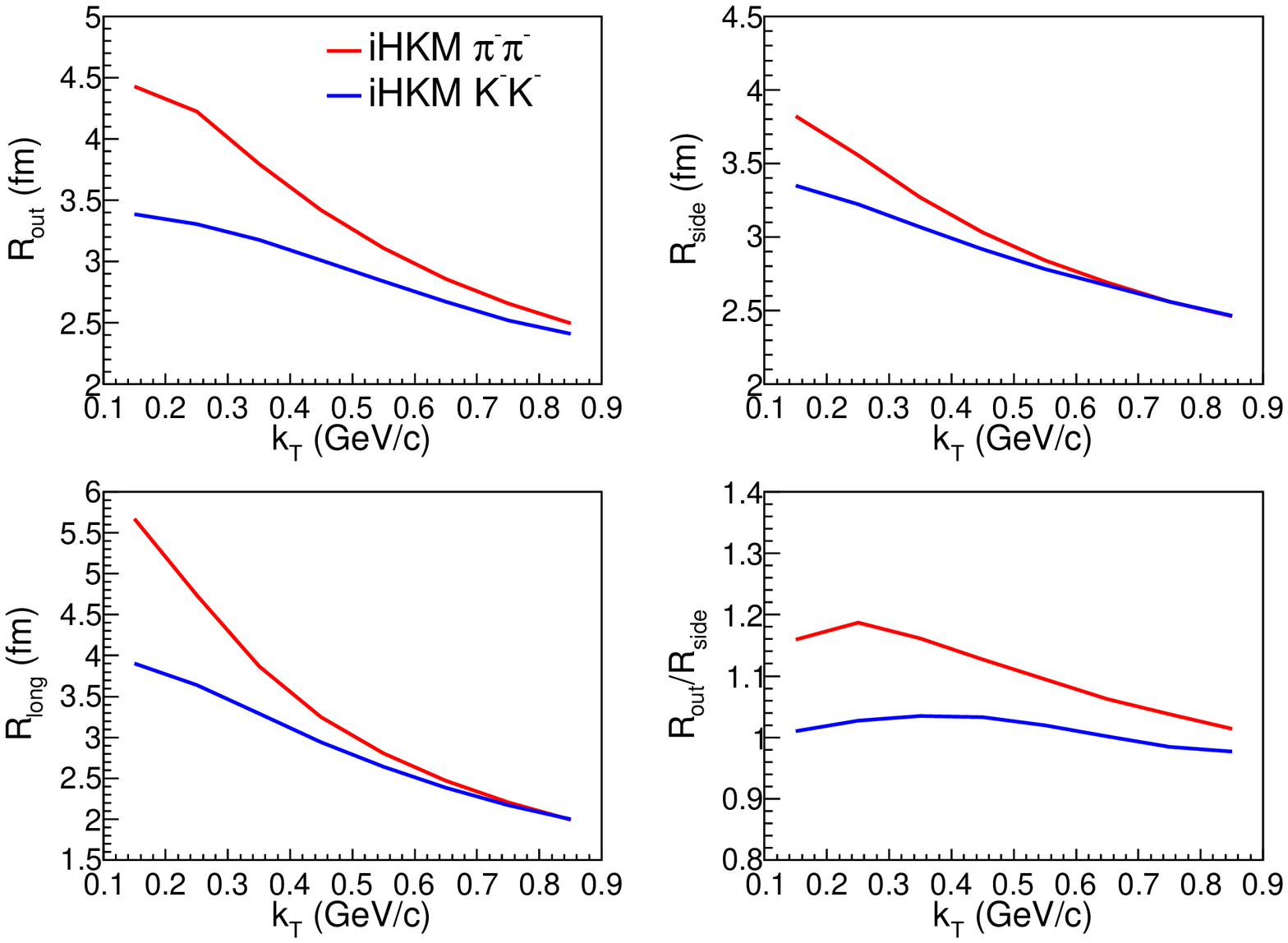}
\caption{The same as in Fig.~\ref{femto_0-10} for the centrality class $c=30-40\%$.}
\label{femto_30-40}
\end{center}
\end{figure}

\begin{figure}
\begin{center}		
\includegraphics[width=0.85\textwidth]{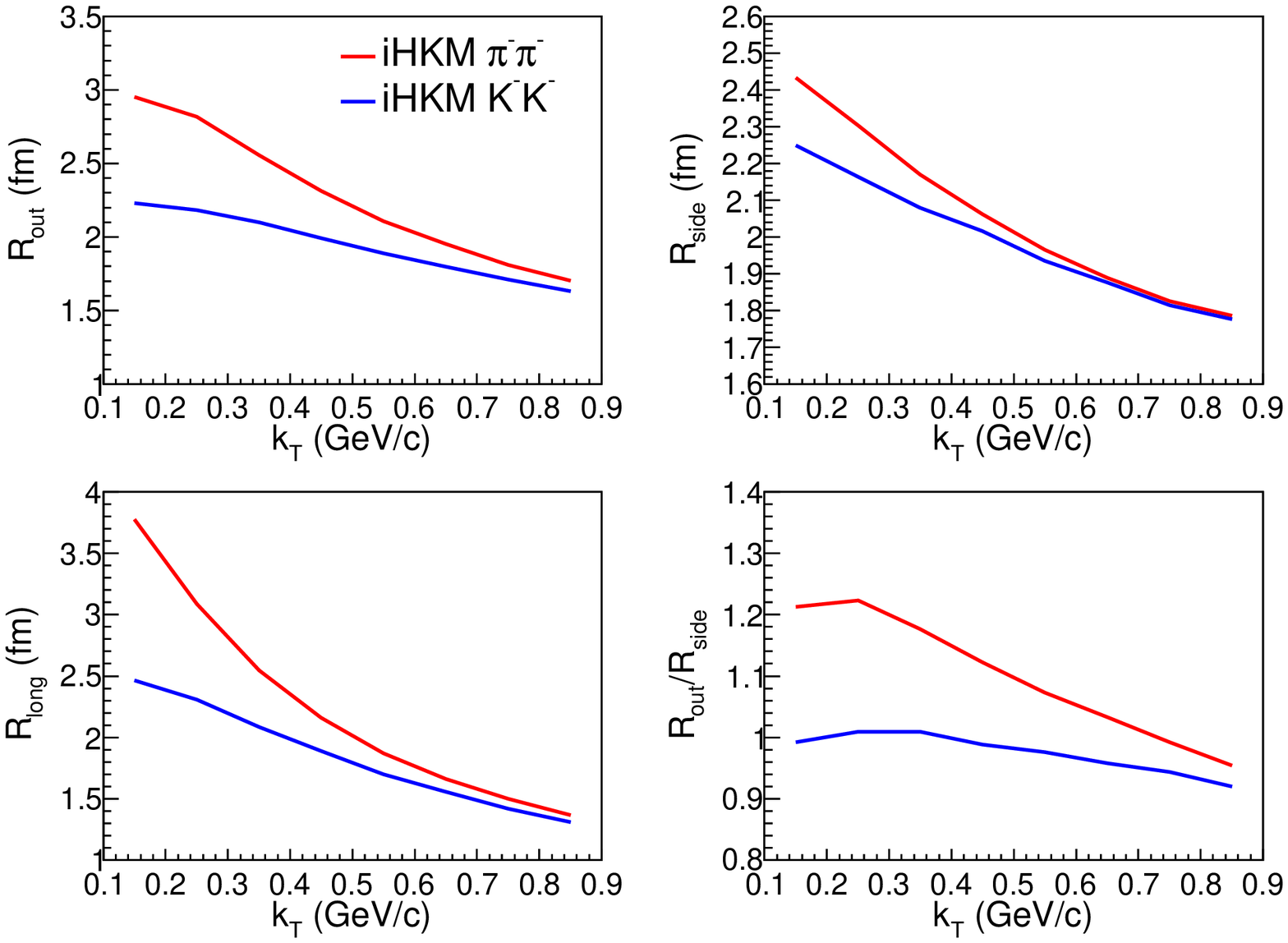}
\caption{The same as in Fig.~\ref{femto_0-10} for the centrality class $c=60-70\%$.}
\label{femto_60-70}
\end{center}
\end{figure}

In Figs.~\ref{femto_0-10}--\ref{femto_60-70} one can find our results on the femtoscopy radii $R_{\mathrm{out}}$, $R_\mathrm{side}$, 
and $R_\mathrm{long}$, extracted from the Gaussian fits to $\pi^{-}\pi^{-}$ and $K^{-}K^{-}$ momentum correlation functions~(CF).
The dependencies of the interferometry radii on the mean pair transverse momentum $k_T$ are presented for the four centrality classes: 
$c=0-10\%$, $c=10-20\%$, $c=30-40\%$, and $c=60-70\%$. 
All the correlation functions are built considering the specially selected particles with $0.15<p_T<1.55$~GeV/$c$ and $|\eta|<1$. 

In the first two figures, the iHKM results for the two centralities, $c=0-10\%$ and $c=10-20\%$, are compared with the experimental data 
on pion and kaon femtoscopy scales from the PHENIX~\cite{phenixfemto} and the STAR~\cite{pionfemto,Grigory} collaborations. 
The STAR data for $KK$ pairs~\cite{Grigory} are preliminary. From the plots it is clear, that iHKM gives a good description of $R_\mathrm{long}(k_T)$ 
dependency both in $\pi\pi$ and in $KK$ case. As for $R_\mathrm{out}$ and $R_\mathrm{side}$ radii, the situation seems to be not so nice.
The model lines for $KK$ look well, but for $\pi\pi$ pairs we see that $R_\mathrm{out}$ values are overestimated, especially for $c=0-10\%$, while
$R_\mathrm{side}$ values are, conversely, underestimated, most noticeably at high $k_T$. As a result, we obtain a rather overestimated
$R_\mathrm{out}/R_\mathrm{side}$ ratio in pion case. 

Also, looking at the figures, one could say that at least $R_\mathrm{side}$ and $R_\mathrm{long}$ 
iHKM curves demonstrate something like scaling between pions and kaons at high $k_T$. For the LHC energies such $k_T$-scaling at $k_T>0.4$~GeV/$c$
for all radii was previously noticed in iHKM simulations~\cite{kaon-our,lhc502} and then for $\sqrt{s_{NN}}=2.76$~TeV Pb+Pb collisions confirmed in the ALICE experimental paper~\cite{alice-scaling}.

In addition, in this paper we apply for the top RHIC energy the method, proposed for the LHC case in~\cite{lifetime} and successfully applied
in~\cite{alice-scaling}, that allows one to extract pion and kaon maximal emission times, $\tau_{\pi}$ and $\tau_K$,
having their $p_T$ spectra and $R_\mathrm{long}(m_T)$ dependencies.

At first we perform a combined fit to pion and kaon transverse momentum spectra, using the analytical formula from~\cite{lifetime}: 
\begin{equation}
p_0 \frac{d^3N}{d^3p} \propto \exp{[-(m_T/T + \alpha)(1-\bar{v}^2_T)^{1/2}]}.
\end{equation}
Here $T$ is the effective temperature, $\alpha$ is a parameter, characterizing the intensity of collective flow (the infinite $\alpha$
means absent flow, while small $\alpha$ values mean strong flow),
and $\bar{v}_T$ is the flow transverse velocity at the saddle point, $\bar{v}_T=k_T/(m_T+\alpha T)$ (see \cite{lifetime} for details).
The spectra fitting is done in $p_T$ range $0.45<p_T<1.0$~GeV/$c$. As a result, we obtain $T=141$~MeV as a common temperature value
for both pions and kaons, and two $\alpha$ values, $\alpha_{\pi}=7.86 \pm 2.11$ and $\alpha_{K}=5.54 \pm 2.61$, for each hadron species respectively. 

After that we use another formula from~\cite{lifetime} to fit kaon and pion $R_\mathrm{long}(m_T)$ dependencies: 
\begin{equation}
R^2_{\mathrm{long}}(m_T)=\tau^2\lambda^2\left(1+\frac{3}{2}\lambda^2\right),
\label{rlongfit}
\end{equation}
where $\lambda$ is connected with the system's homogeneity length in longitudinal direction $\lambda_l$, namely 
$\lambda^2=(\lambda_l/\tau)^2=T/m_T\cdot(1-\bar{v}^2_T)^{1/2}$, and $\tau$ is the corresponding maximal emission time.
Our values of $R_\mathrm{long}$ radii were obtained from Gaussian fits to the corresponding correlation functions  
in the particle momentum difference $q$ interval $|q|<0.3$~GeV/$c$.

\begin{figure}
\begin{center}
\includegraphics[width=0.7\textwidth]{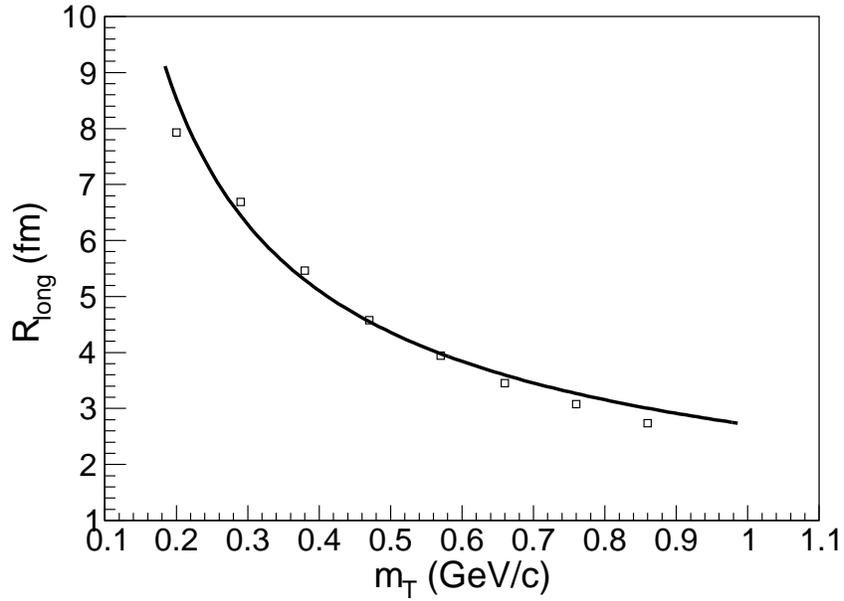}
\caption{The $R_{long}$ dependency on pair $m_{T}$ for the negatively charged pion pairs together with the fit to it according to formula (\ref{rlongfit}).}
\label{pilong}
\end{center}
\end{figure}

\begin{figure}
\begin{center}
\includegraphics[width=0.7\textwidth]{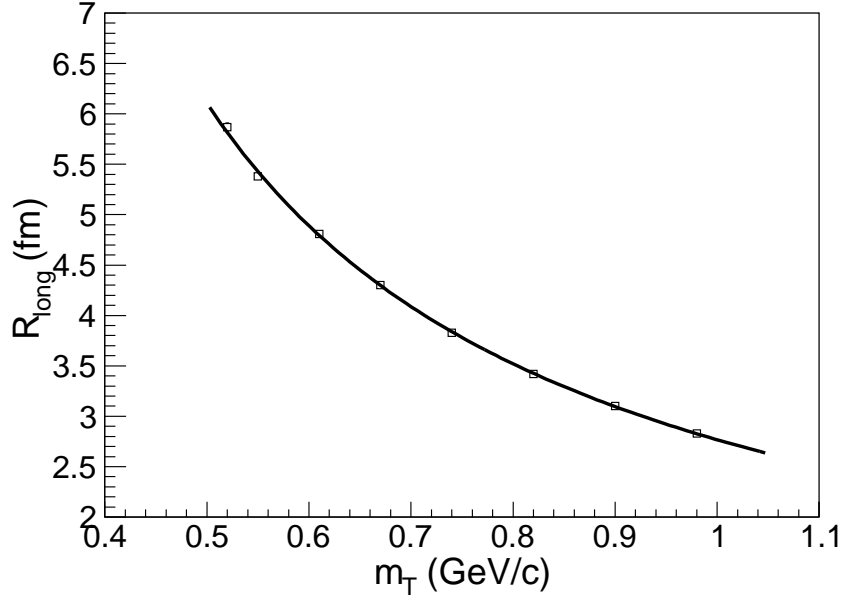}
\caption{The same as in Fig.~\ref{pilong} for negatively charged kaon pairs.}
\label{klong}
\end{center}
\end{figure}

Then, having constrained the $T$ and $\alpha$ parameters according to the results of combined $p_T$-spectra fitting, 
we extract the desired maximal emission time $\tau_{\pi}=7.12\pm 0.01$~fm/$c$ for pions from the fit to pion
$R_\mathrm{long}(m_T)$ dependency using the formula (\ref{rlongfit}).
In order to obtain the $\tau$ value for kaons, similarly to the LHC case, described in~\cite{lifetime},
we have to set $\alpha$ parameter free at fitting kaon $R_\mathrm{long}$ points.
Eventually, we obtain maximal emission time $\tau_{K}=9.71 \pm 0.02$~fm/$c$, and the kaon $\alpha$ value $\alpha_{K}=0.12 \pm 0.02$.
One can see both iHKM $R_\mathrm{long}(m_T)$ dependencies together with fits to them in Figs.~\ref{pilong}, \ref{klong}.

\begin{figure}
\begin{center}
\includegraphics[width=0.98\textwidth]{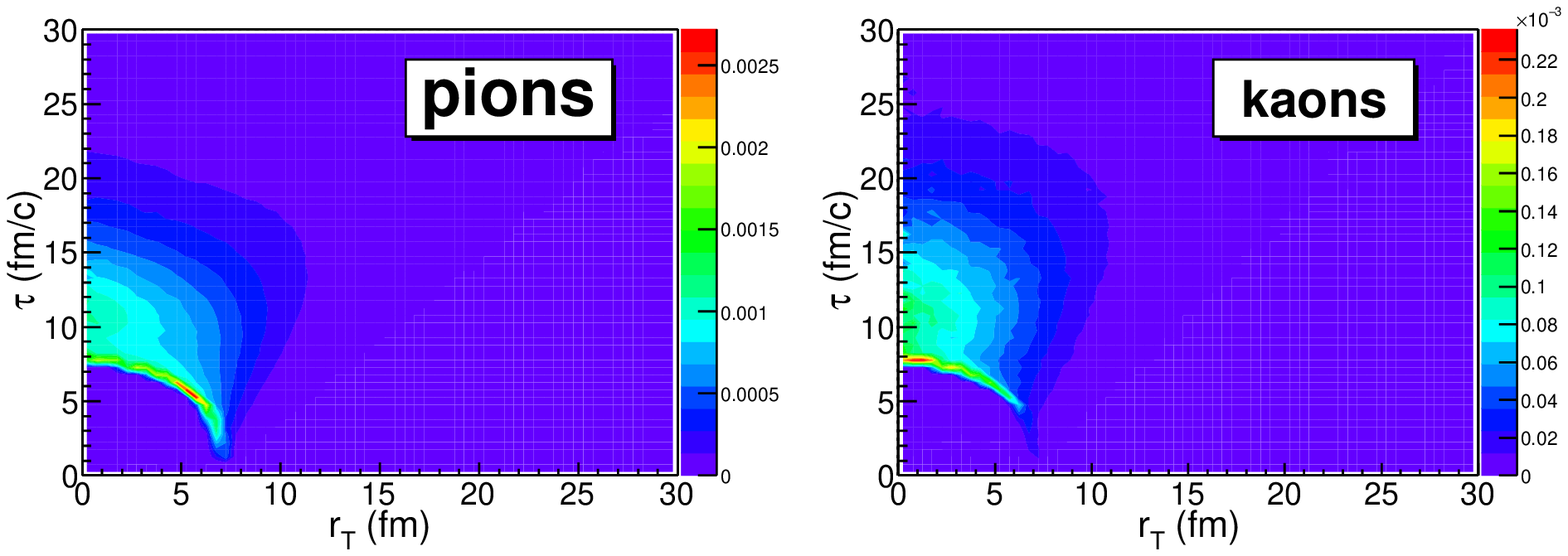}
\caption{The pion and kaon emission functions $g(\tau,r_T,p_T)$ [fm$^{-3}$], averaged over complementary variables, obtained from iHKM for
the centrality class $c=0-10\%$.
Particles with $0.2<p_T<0.3$~GeV/$c$ and $|\eta|<1$ were chosen for the analysis.}
\label{emiss}
\end{center}
\end{figure}

Finally, in Fig.~\ref{emiss} we present the iHKM pion and kaon averaged emission functions $g(\tau,r_T,p_T)$, which reveal the space-time picture
of radiation of these particles. The maximal emission times can be approximately found from these plots, if one attributes
some $\tau$ values to the regions, where each $g(\tau,r_T,p_T)$ has maximum. As it is readily seen, maximal emission time values obtained
in this way are close to those accurately extracted from fits. More detailed analysis, provided in Ref.~\cite{kstar}, shows that the reason 
for a larger time of maximal emission, obtained for kaons, as compared with pions, is in intensive decays and recombinations of $K^*$ mesons
(having life-time near 4~fm/$c$), which take place at the afterburner stage of the collision. 
This effect was also found in the ALICE experimental analysis~\cite{alice-scaling}. 

\section{Conclusions}
The integrated hydrokinetic model showed itself not less successful in describing the variety of bulk observables in Au+Au collisions at the top 
RHIC energy, than in Pb+Pb ones at the available LHC energies. After adjusting the main model parameters, namely the maximal initial energy 
density $\epsilon_0(\tau_0)$ and the binary collision
contribution to the initial transverse energy-density profile $\alpha$, the iHKM allowed to describe simultaneously the experimental particle yields and 
their ratios, $p_T$ spectra for pions, kaons and (anti)protons, and $v_2$ coefficients for all charged particles.

As for the femtoscopy scales, they are described in the model well for kaon pairs, and in the pion case we observe some overestimation of
$R_\mathrm{out}$ and underestimation of $R_\mathrm{side}$ radius, more pronounced at high pair $k_T$. 

The longitudinal radii $R_\mathrm{long}$ perfectly describe the experimental data at different centralities.
So, they are used to extract the times of the maximal emission for pions and kaons according to the procedure, proposed for the LHC case 
in the paper~\cite{lifetime}. We found that both corresponding times are about 2~fm/$c$ less than at the LHC energy $\sqrt{s_{NN}}=2.76$~TeV. 
Similarly as at the LHC, the maximal emission time for kaons is larger than for pions. And again, we explain the latter fact by the intensive decays 
and recombinations of $K^*$ resonances at the final stage of the collision.

\begin{acknowledgments}
The research was carried out within the scope of the EUREA: European Research Network ``Heavy ions at ultrarelativistic energies'' and corresponding Agreement 
with the National Academy of Sciences (NAS) of Ukraine. The work is partially supported by
the NAS of Ukraine Targeted research program ``Fundamental research on high-energy physics and nuclear physics (international cooperation)''.
The publication contains the results of studies conducted by President’s of Ukraine grant for competitive projects (project number F75/219-2018) of the State Fund for Fundamental Research.  
\end{acknowledgments}

\end{document}